\documentclass[conference,9pt]{IEEEtran}
\IEEEoverridecommandlockouts
\usepackage{cite}
\usepackage[acronym]{glossaries}
\usepackage{subcaption}
\usepackage{amsmath,amssymb,amsfonts}
\usepackage{algorithmic}
\usepackage{graphicx}
\usepackage{textcomp}
\usepackage[table,svgnames]{xcolor}
\usepackage{setspace}

\usepackage{threeparttable}
\usepackage{tabulary}
\usepackage{booktabs}
\usepackage{adjustbox}
\usepackage{siunitx}
\def\BibTeX{{\rm B\kern-.05em{\sc i\kern-.025em b}\kern-.08em
    T\kern-.1667em\lower.7ex\hbox{E}\kern-.125emX}}
\begin{document}

\setstretch{0.82}

\bstctlcite{IEEEexample:BSTcontrol}

\newacronym[plural=TNNs, firstplural={Ternary Neural Networks (TNNs)}]{tnn}{TNN}{Ternary Neural Network}
\newacronym{cutie}{CUTIE}{Completely Unrolled Ternary Inference Engine}
\newacronym{soc}{SoC}{System-on-Chip}
\newacronym{dnn}{DNN}{Deep Neural Network}
\newacronym{dvs}{DVS}{Dynamic Vision Sensor}
\newacronym{dvss}{DVSs}{Dynamic Vision Sensors}
\newacronym{iot}{IoT}{Internet of Things}
\newacronym[plural=SNNs, firstplural={Spiking Neural Networks (SNNs)}]{snn}{SNN}{Spiking Neural Network}

\title{Kraken: A Direct Event/Frame-Based Multi-sensor Fusion SoC for Ultra-Efficient Visual Processing in Nano-UAVs}
\author{\IEEEauthorblockN{Alfio Di Mauro\IEEEauthorrefmark{2}\IEEEauthorrefmark{4}, Moritz Scherer\IEEEauthorrefmark{2}\IEEEauthorrefmark{4}, Davide Rossi\IEEEauthorrefmark{3}, Luca Benini\IEEEauthorrefmark{2}\IEEEauthorrefmark{3}}
\IEEEauthorblockA{\IEEEauthorrefmark{2} ETH Z\"{u}rich, Switzerland, \IEEEauthorrefmark{3} University of Bologna, Italy  \IEEEauthorrefmark{4} equal contribution}}%
\vspace{-1cm}
\maketitle
\section{Introduction}
\vspace{-0.05cm}
Small-size unmanned aerial vehicles (UAV) have the potential to dramatically increase safety and reduce cost in applications like critical infrastructure maintenance and post-disaster search and rescue\cite{Skydio}. Many scenarios require UAVs to shrink toward nano and pico-size form factors \cite{pico-drone}.
The key open challenge to achieve true autonomy on Nano-UAVs is to run complex visual tasks like object detection, tracking, navigation and obstacle avoidance fully on board, at high speed and robustness, under tight payload and power constraints. 
With the Kraken \gls{soc}, fabricated in \SI{22}{\nano\meter} FDX technology (Fig. \ref{tab:kraken_details},\ref{fig:dieshot}), we demonstrate a multi-visual-sensor capability exploiting both event-based and BW/RGB imagers, combining their output for multi-functional visual tasks previously impossible on a single low-power chip for Nano-UAVs\cite{Skydio}.
Kraken is an ultra-low-power, heterogeneous \gls{soc} architecture integrating three acceleration engines and a vast set of peripherals to enable efficient interfacing with standard frame-based sensors and novel event-based \gls{dvss}\cite{dvs}. 
Kraken enables highly sparse event-driven sub-\SI{}{\micro \joule}/inf \gls{snn} inference on a dedicated neuromorphic energy-proportional accelerator. Moreover, it can perform frame-based inference by combining a \SI{1.8}{TOp\per\second\per\watt} 8-cores RISC-V processor cluster with mixed-precision \gls{dnn} extensions with a \SI{1036}{TOp\per\second\per\watt} \gls{tnn} accelerator.

\vspace{-0.2cm}
\section{Architecture}
The Kraken SoC (Fig. \ref{fig:syslevel}) is built around a 32bit fabric controller (FC) RISC-V core, it hosts 1MiB of scratchpad SRAM memory (L2) and standard peripherals: 4 QSPI, 4 I2C, 2 UART and 48 GPIOs. The FC can offload compute intensive kernels to three programmable, power-gateable accelerators.
\subsubsection{Sparse Neural Engine (SNE)}
It targets spiking convolutional neural network (SCNN) inference with 4bit 3x3 kernels and 8bit leaky-integrate and fire (LIF) neuron states. SNE exploits an explicit coordinate list (COO) data representation to efficiently transform unstructured spatio/temporal sparse event computation, which is difficult to perform efficiently, into SNE ``dense'' computational bursts.
SNE hosts eight 8KiB LIF neuron state memories and a dedicated 9.2KB weight buffer. 
\subsubsection{Completely Unrolled Ternary Inference Engine (CUTIE)}
It is a \gls{tnn} accelerator designed to maximize energy efficiency by minimizing data movement during inference. This is achieved by keeping all ternary weights on-chip (in 1.6 bits/weight compressed format), fully spatially unrolling the ternary multiplications, and the multi-bit accumulation required to compute an output channel activation followed by per-channel normalization, non-linearity, and thresholding. CUTIE achieves one output activation element per cycle per output channel throughput. 
The CUTIE instance in Kraken supports 96 parallel output channels, 158kB, and 117kB memories for feature map and weight storage, respectively.
\subsubsection{Parallel ultra-low power cluster (PULP)}
It hosts 8 RISC-V cores sharing a single-cycle 128kiB L1 scratchpad memory. The cores feature dedicated extensions for energy-efficient digital signal processing such as hardware loops, multiply-accumulate with concurrent load (\textit{MAC-LD}) and multi-precision (\textit{fp32/fp16/fp16-brain}) floating point. The Cluster also supports SIMD (\textit{int8/int4/int2}) widening dot-product operations, as well as all their mixed-precision combinations, thanks to a status-based RISC-V ISA extension. 

\vspace{-0.2cm}
\section{Applications and results}
\begin{figure*}
\vspace{1cm}
\begin{minipage}[t!]{0.5\linewidth}
  \begin{center}
    \includegraphics[width=\linewidth]{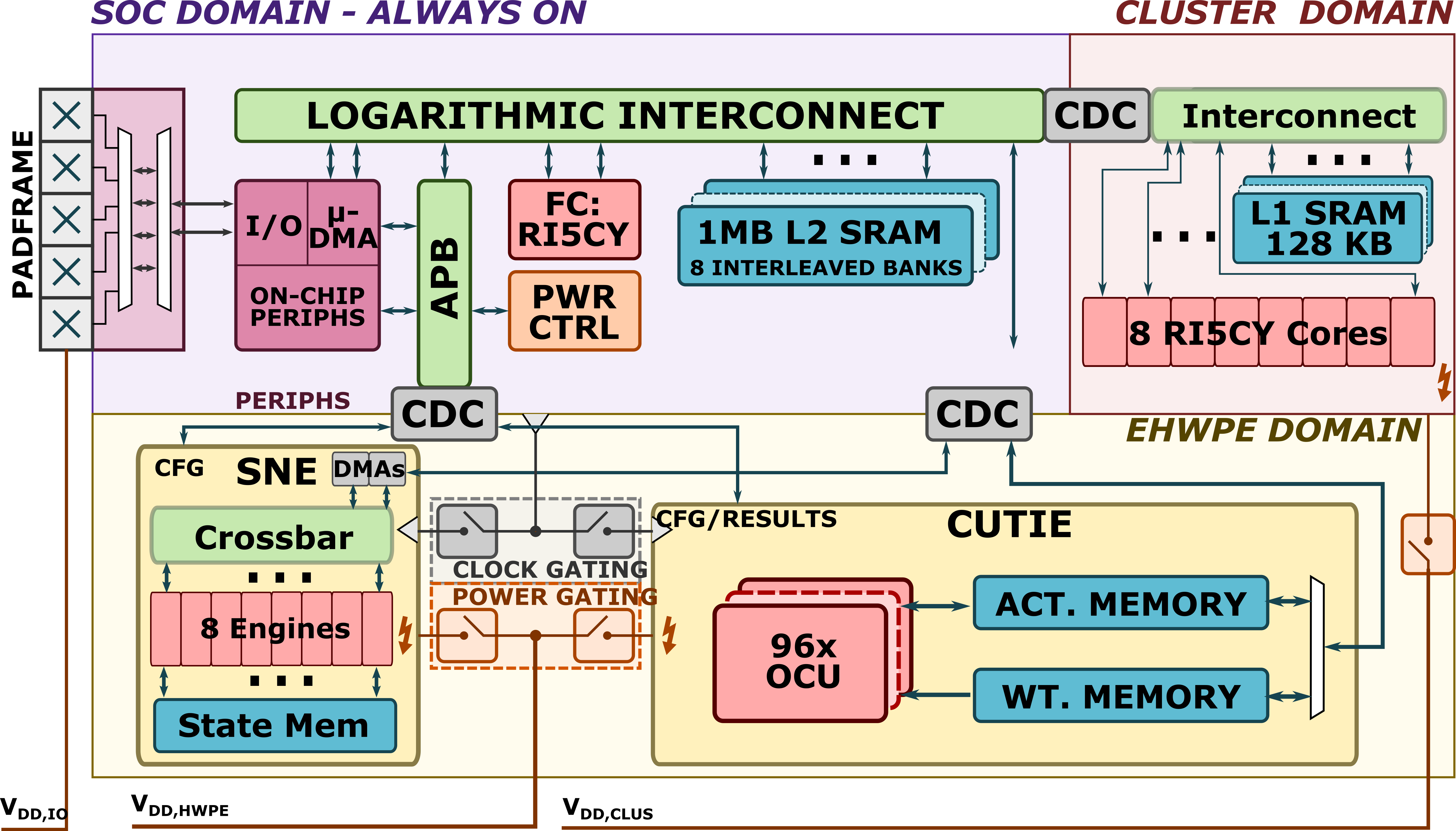}
    \caption{System level overview showing the three main computation engines and their integration into the SoC.}
    \label{fig:syslevel}
  \end{center}
  \end{minipage}\hfill%
\begin{minipage}[t!]{0.45\linewidth}
  \begin{center}
  \includegraphics[width=\linewidth]{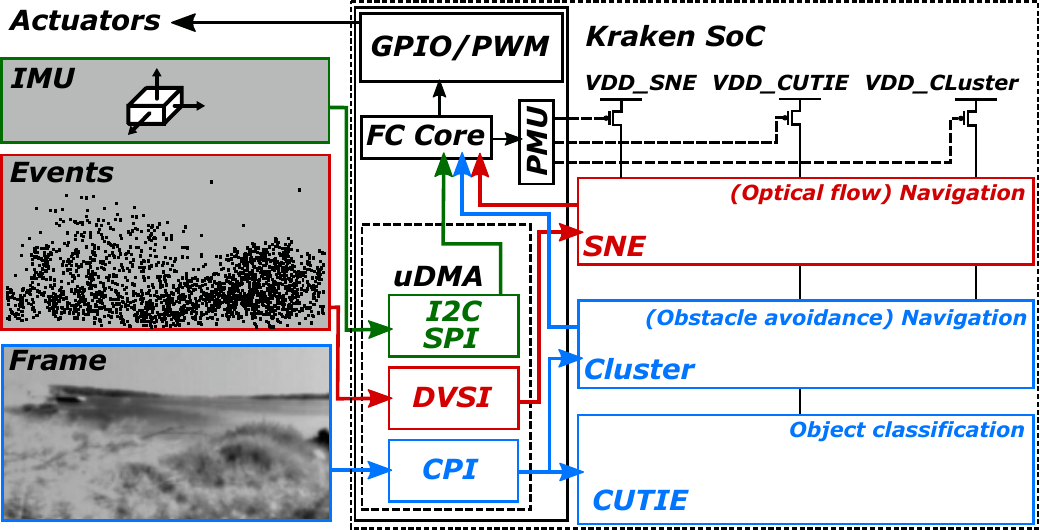}\\
    \caption{Description of the application scenario. The two computation subsystems are shown in red and blue respectively.}
    \label{fig:application_drone}
    \end{center}
\end{minipage}
\vspace{1cm}
\begin{minipage}[t!]{0.35\linewidth}
  \begin{center}
    \includegraphics[width=0.85\linewidth]{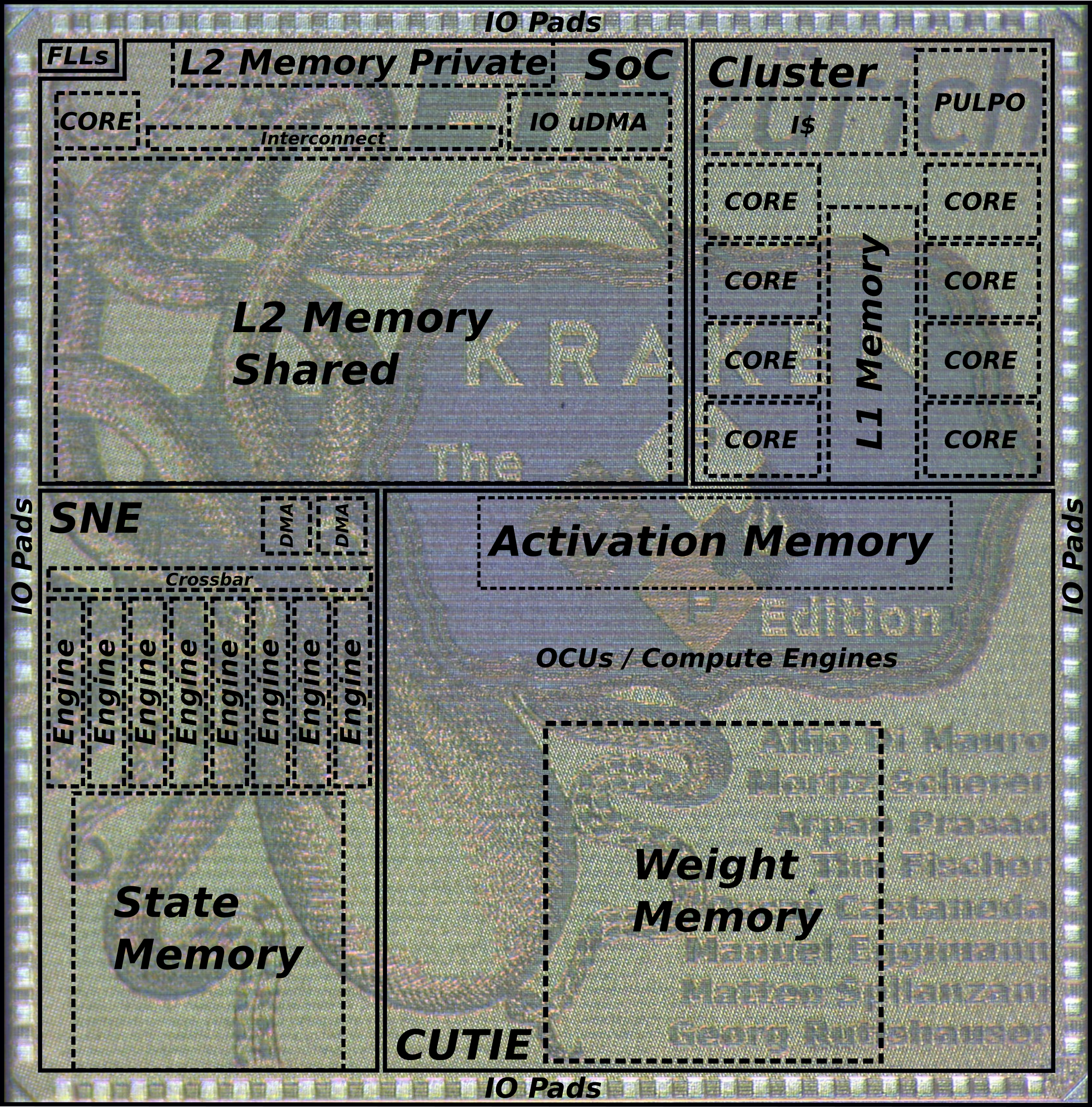}
    \caption{Die micrograph of the Kraken SoC showing the different power domains.}
    \label{fig:dieshot}
  \end{center}
\end{minipage}%
\hfill
\begin{minipage}[t!]{.31\linewidth}
  \begin{center}
    \includegraphics[width=\linewidth]{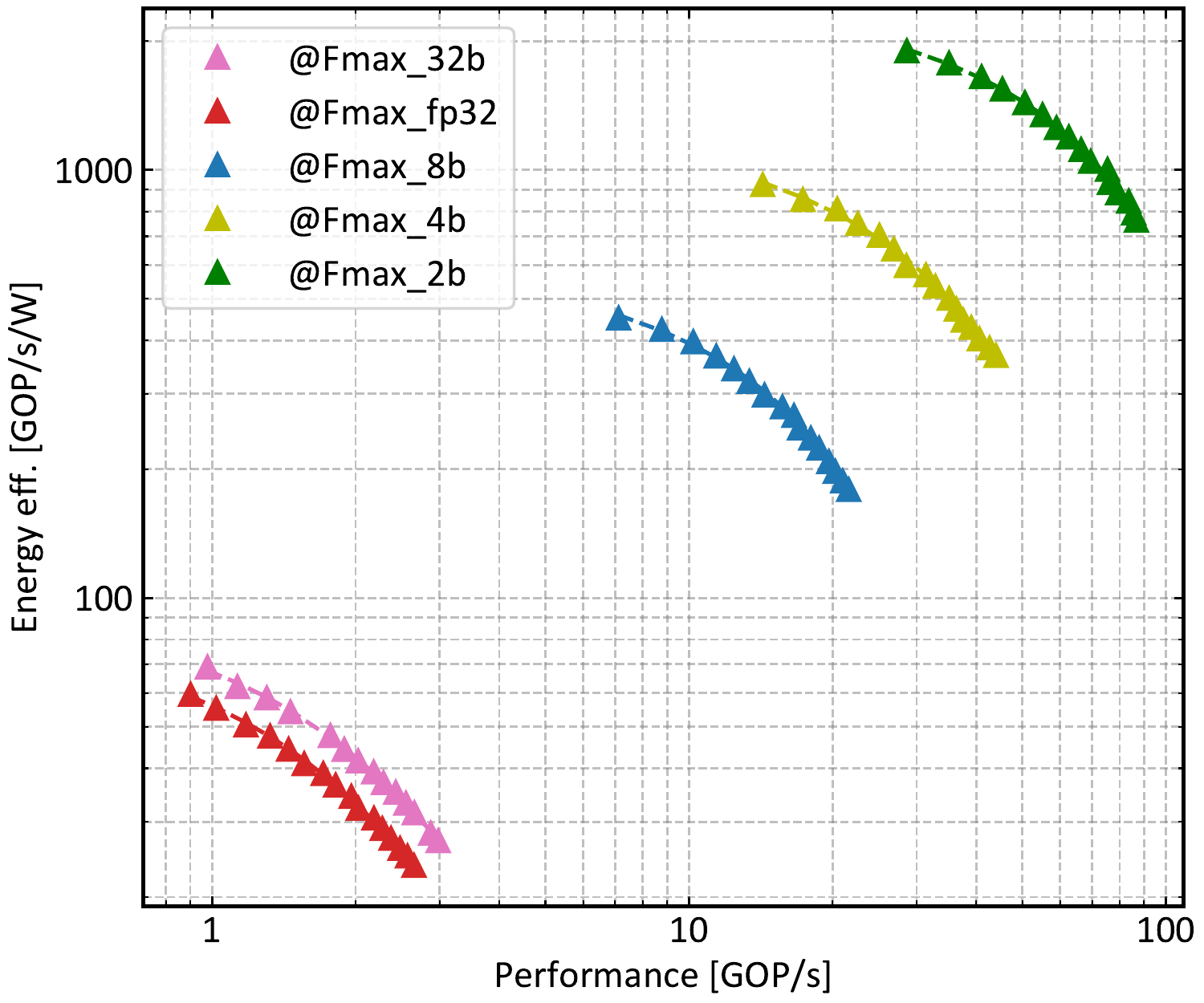}
    \vspace{-0.3cm}
    \caption{Energy efficiency of the PULP cluster for different bit precision. unless ``fp'' specified, integer arithmetic is indented. SIMD instructions are used below 32bit precision.}
    \label{fig:cluster}
  \end{center}
\end{minipage}
\hfill%
\begin{minipage}[t!]{0.33\linewidth}
    \centering
    \resizebox{0.75\linewidth}{!}{
    \begin{tabular}{@{}ll@{}}
    \toprule
    Technology & GF \SI{22}{\nano\meter} FDX \\ \midrule
    Chip area & \SI{9}{\milli\meter\squared} \\ \midrule
    L2 memory (SRAM) & 1MiB  \\ \midrule
    L1 Memory (SRAM) & 128kiB \\ \midrule
    VDD Range & \SIrange{0.5}{0.8}{\volt} \\ \midrule
    Cluster Max. Frequency & \SI{330}{\mega\hertz} \\ \midrule
    EHWPE Max. Frequency & \SI{330}{\mega\hertz}\\ \midrule
    FC Max. Frequency & \SI{330}{\mega\hertz} \\ \midrule
    Power Range & \SIrange{2}{300}{\milli\watt} \\ \bottomrule
    \end{tabular}%
    }
    \caption{Physical implementation details for the kraken chip. Maximum frequency has been measured on 32bit floating point matrix-to-matrix multiplication benchmarks for both FC and cluster. For CUTIE and SNE, CNN workloads have been used.}
    \label{tab:kraken_details}
\end{minipage}%
\vspace{-0.4cm}
\begin{minipage}[t!]{.49\linewidth}
    \includegraphics[width=\linewidth]{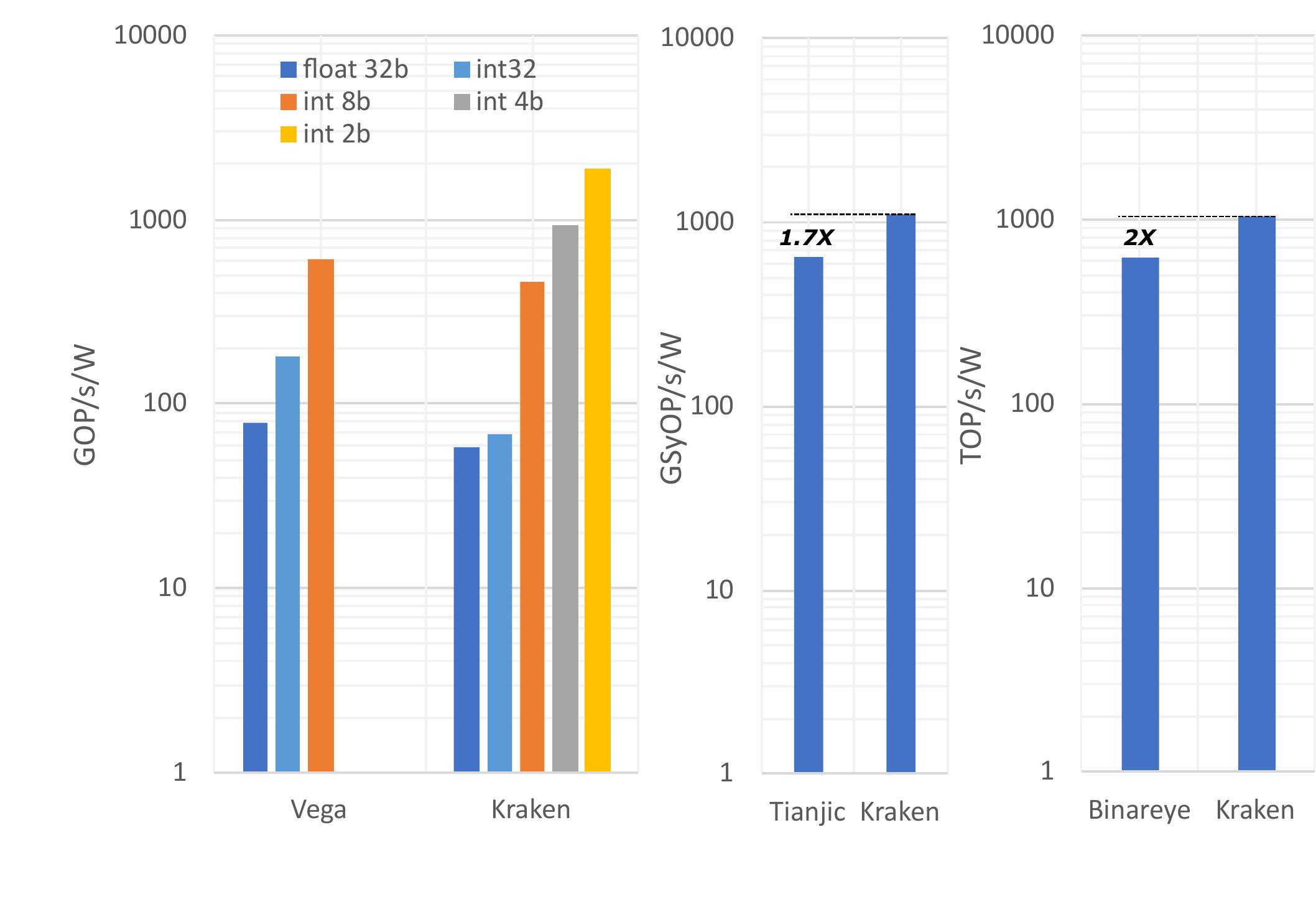}
    \caption{Comparison of the energy efficiency of the cluster, SNN accelerator and TNN accelerator with state-of-the-art designs. In the case of PULP, 2 N-bit OP = 1 N-bit MAC. For CUTIE, 2 ternary OP = 1 ternary MAC. In the case of SNE, 1 SOP = 1 4b-ADD + 1 8b-MUL + 1 8b-COMPARE
  }
    \label{fig:efficiency}
\end{minipage}
\hfill%
\begin{minipage}[t!]{.49\linewidth}
  \begin{center}
    \includegraphics[width=\linewidth]{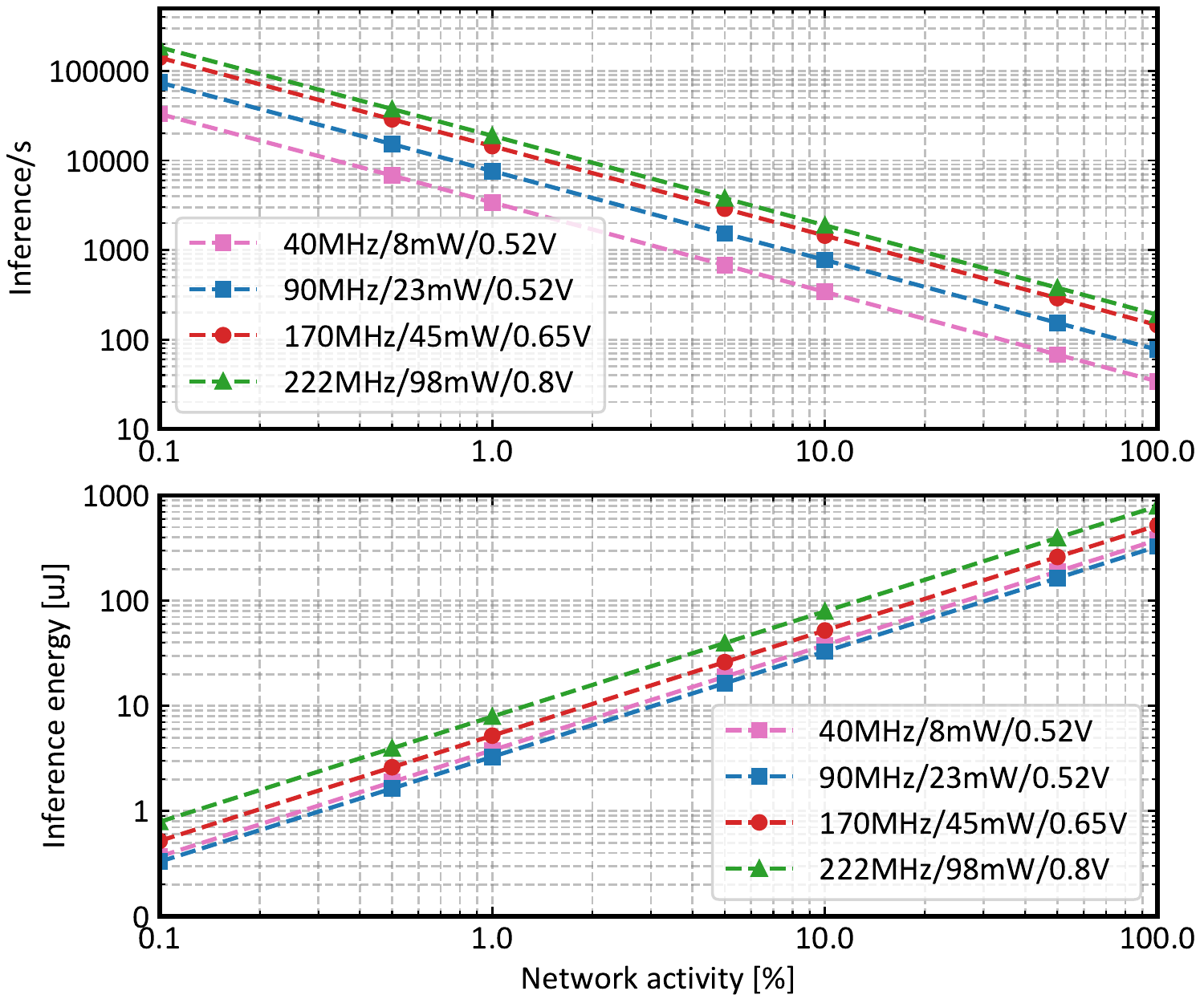}
    \caption{SNE inference per second versus the DVS network activity (Top). SNE inference energy versus DVS network activity (Bottom).}
    \label{fig:sne}
  \end{center}
\end{minipage}\hfill%
\end{figure*}

We focus on a nano-UAV workload composed of three key visual tasks (Fig. \ref{fig:application_drone}). The SNE subsystem assists navigation by providing optical flow reconstruction from DVS events (produced by a DVS132s 128x132 by IniVation\footnote{https://inivation.com}), while PULP and CUTIE execute obstacle avoidance and target object detection, respectively, on BW images (produced by the HM01B0 320X240 imager from HIMAX\footnote{https://www.himax.com.tw}).
We present post-silicon measurements to assess the energy consumption of the aforementioned tasks.
SNE can compute per pixel optical flow with a 4bit quantized, 4-layer, Convolutional Spiking Neural Network (CSNN), low-memory footprint, LIF-FireNet proposed in \cite{NEURIPS2021_39d4b545}. During inference, the SNE consumes 98mW at 222MHz, 0.8V. At this frequency, SNE can perform 20800 inf/s at low (1\%) network activity, and 1019 inf/s at the 20\% average activity (Fig. \ref{fig:sne}).
In parallel, CUTIE can perform object classification at more than 10000 inf/s on a ternary CIFAR10 network derived from\cite{Moons2018}, in a power envelope of 110mW, at 0.8V, when clocked at 330MHz.
The PULP cluster performs a navigation and obstacle avoidance task based on an 8bit quantized DroNet network presented in\cite{pico-drone}. The network can be executed at a rate of 28inf/s when PULP is running at 330MHz, 0.8V, in an 80mW power envelope.

SNE energy efficiency has been benchmarked against SoA on a 6-layer CSNN network presenting similar complexity and memory footprint as the LIF-FireNet one. On the standard event-based IBM-DVSGesture, our design can achieve SoA 92\% accuracy while showing an energy efficiency that outperforms the state-of-the-art by 1.7$\times$ \cite{Tianjic}.
CUTIE has been benchmarked on a ternarized version of the binary network reported in \cite{Moons2018}, performing object detection on the CIFAR10 data set. 
It achieves 2\% better accuracy than  \cite{Moons2018} and energy efficiency of \SI{1036}{TOp\per\second\per\watt}, outperforming the state-of-the-art by 2$\times$\cite{Moons2018}.
To benchmark the PULP cluster energy efficiency against a similar SoA RISC-V cluster\cite{Rossi2022}, we executed standalone convolutional layer patches that are representative of multi-precision DNN inference.
Compared to Vega\cite{Rossi2022}, on the same workload, Kraken show 1.66$\times$ higher throughput at the same frequency, thanks to the \textit{MAC-LD} instruction, allowing to achieve a peak throughput of 0.98 mac/cycle/core. 
In terms of energy efficiency, the SIMD operations for highly quantized inference enable Kraken to achieve more than 2.6$\times$ better energy efficiency on 4-b and 2-b convolutions (Fig. \ref{fig:cluster}).
Kraken engines' energy efficiency is summarized in Fig. \ref{fig:efficiency} and compared against each SoA counterpart.
By combining SNE, CUTIE, and PULP with the microcontroller FC subsystem, Kraken's heterogeneous SoC architecture can concurrently execute all visual tasks required for autonomous navigation on Nano-UAVs.
\vspace{-0.1cm}

\bibliography{./bibliography}

\newpage

\end{document}